\documentclass[         %
aps,                    
prd,                    
showpacs,               
superscriptaddress,    
nofootinbib,            
twocolumn,             %
showkeys,               %
preprintnumbers,        %
floatfix               
]{revtex4}               
\usepackage{graphicx,amsmath}
\usepackage[justification=RaggedRight]{caption}
\usepackage{enumerate}
\usepackage{xcolor}
 \usepackage{float}
\usepackage{subfigure}

\begin{document}
\title{Matter-Neutrino Resonance Above Merging Compact Objects
}
\author{Annelise~Malkus}
\email{acmalkus@ncsu.edu}
\affiliation{Department of Physics, North Carolina State University, Raleigh, NC 27695, USA}

\author{Alexander Friedland}
\email{friedland@lanl.gov}
\affiliation{Theoretical Division, T-2, MS B285, Los Alamos National Laboratory, Los Alamos, NM 87545-0285, USA}

\author{Gail~C.~McLaughlin}
\email{gcmclaug@ncsu.edu}
\affiliation{Department of Physics, North Carolina State University, Raleigh, NC 27695, USA}

\date{March 30, 2014}
\begin{abstract}
Accretion disks arising from neutron star- neutron star mergers or black hole- neutron star mergers produce large numbers of neutrinos and antineutrinos.
In contrast to other astrophysical scenarios, like supernovae, in mergers the antineutrinos outnumber the neutrinos.
This antineutrino dominance gives neutrinos from merger disks the opportunity to exhibit new oscillation physics, specifically a matter-neutrino resonance. 
We explore this resonance, finding that consequences can be a large transition of $\nu_e$ to other flavors, while the $\bar{\nu}_e$s return to their initial state.  We present numerical calculations of neutrinos from merger disks and compare with a single energy model.
We explain both the basic features and the conditions for a transition.
\end{abstract}
\medskip
\maketitle

\preprint{LA-UR-14-21712}
\preprint{CETUP2013-020}

The merger of two neutron stars, or a neutron star and a black hole, forms a black hole accretion disk.
These mergers are fascinating for many reasons: as home to large numbers of neutrinos, to dense matter physics \cite{lattimer:2012}, jets \cite{Blinnikov} and gravitational waves \cite{harry,duez,baumgarte}, as well as r-process \cite{lattimer,Surman:2008qf,Rosswog:2013kqa} and other types of nucleosynthesis \cite{Pruet:2003yn,Surman:2011aa,Surman:2013sya}.
The neutrinos play a significant role in disk dynamics \cite{MacFadyen:1998vz}, jet production \cite{Ruffert:1998qg,Oechslin:2005mw}, and wind-type nucleosynthesis \cite{Surman:2003qt,Surman:2004sy,Surman:2005kf}.
Neutrinos, however, can transform away from their flavor composition at emission and this can have 
important consequences, particularly for the outcome of the wind-type nucleosynthesis \cite{Qian:1994wh,Pastor:2002we,Balantekin:2004ug,Duan:2010af}.
Below,
we examine the neutrino flavor transformation physics in merger disk environments and describe a phenomenon, which we call a matter-neutrino resonance (MNR) transition.%

Neutrino physics has changed dramatically in the past few years.
Calculations that take into account coherent neutrino self-interactions in conditions typical of core collapse supernovae have shown that the neutrinos exhibit significant -- and highly nontrivial -- flavor transformations \cite{Duan:2006jv,Duan:2006an,Balantekin:2006tg,Hannestad:2006nj,Raffelt:2007cb,Duan:2007mv,Fogli:2007bk,Duan:2007sh,Dasgupta:2008cd,Gava:2008rp,Dasgupta:2009mg,Friedland:2010sc,Duan:2010bf,Choubey:2010up,Galais:2011gh,Pehlivan:2011hp}.
While the high density of neutrinos near emission leads to energy synchronized neutrino flavor evolution there \cite{Pastor:2001iu,Friedland:2006ke}, 
a remarkable phenomenon occurs further out:
As the neutrino self interaction potential drops toward the vacuum scale, $\Delta = \delta m^2 / (2 E)$, where $\delta m^2$ is the mass splitting of the neutrinos and $E$ is their energy \cite{Hannestad:2006nj,Duan:2007mv,Duan:2007sh,Dasgupta:2008cd}, both neutrinos and antineutrinos can oscillate, eventually forming spectral splits.

Mergers present an oscillation environment not possible in settings studied earlier.
Since the material in compact object merger disks begins heavily neutron rich, as it heats it tends to leptonize, i.e. emit more antineutrinos $\bar\nu_{e}$ than neutrinos  $\nu_{e}$ \cite{Ruffert:1996by,Surman:2008qf,Deaton:2013sla, Kiuchi:2012mk}.
The excess of $\bar\nu_{e}$ over $\nu_{e}$ means that the neutrino self-interaction potential has \emph{opposite sign} to the matter potential.  
When the two potentials cancel an MNR transition can occur.
This transition is a collective phenomenon that is physically distinct from the supernova collective transformations outlined above: MNR can occur deep in the would-be synchronization region, in which both matter and self-interaction potentials are much bigger than the vacuum scale. MNR breaks the synchronization between neutrinos and antineutrinos, allowing neutrinos to completely change flavor, while returning antineutrinos to their original state.  The MNR phenomenon is also distinct from the ordinary MSW effect, as follows already from the fact that MNR occurs for either mass hierarchy.

A transition at such a resonance point was observed in  \cite{Malkus:2012ts} in the context of disks from stellar collapse. While the resonance condition is similar to that from stellar collapse disks, the resonance in merger disks causes novel behavior.
The transformation occurs close to the emission surface, which may have significant implications for wind type nucleosynthesis.%
In this paper we explain the  physics of this transition using a single-energy model and provide an analytic formula to describe the resulting transition.  
After elucidating the mechanism of MNR, we present calculations of transformation above merger disks.

{\it Matter-Neutrino Resonance:}
The simplest system exhibiting MNR is a two-flavor model of neutrinos and antineutrinos of a single energy.
Like any $SU(2)$ system, this problem can be easily recast in the language of the Neutrino Flavor Isospin (NFIS) \cite{Duan:2005cp} formalism,
named for the flavor isospin vectors, ${\bf s}(\bar{\bf s})$, such that, {\it e.g.}, a neutrino state $|\nu\rangle$ maps into a vector ${\bf s} = \langle \nu|{\vec\sigma}/2|\nu\rangle$.
These vectors have length $1/2$, and the third($\hat{\bf z}$) components are $s_z = P_{\nu_e} - 1/2 $($\bar{s}_z = -P_{\bar{\nu}_e} + 1/2$),
where $P_{\nu_e}$ ($P_{\bar{\nu}_e}$) is the survival probability for an electron (anti)neutrino (see \cite{Duan:2006an} for the full definitions).
In their entirety, the evolution equations are,
\begin{align}
\frac{\partial {\bf s}}{\partial l} &= {\bf s}\times\left[ \Delta {\bf H}_V + V_e\hat{\bf z} + 2\mu_\nu ({\bf s} + \alpha {\bf \bar{s}}) \right]\label{eq:NFIS1},
\end{align}
\begin{align}
\frac{\partial{\bf \bar{s}}}{\partial l} &= {\bf \bar{s}}\times\left[ -\Delta {\bf H}_V + V_e \hat{\bf z} + 2\mu_\nu ({\bf s} + \alpha {\bf \bar{s}}) \right],
\label{eq:NFIS2} 
\end{align}
where ${\bf H }_V = (-\sin 2 \theta_V, 0, \cos 2 \theta_V)$ depends on the vacuum mixing angle, $\theta_V$, $\mu_\nu$ is the neutrino-neutrino interaction strength, and $\alpha$ is the ratio of the unoscillated $\bar\nu_{e}$ and $\nu_{e}$ fluxes.
We use $\theta_V=0.15$, which is consistent with the recommended value of $\theta_{13}$ \cite{An:2013zwz}.
The sign of $\Delta$ determines the hierarchy.
We assume that the neutrinos start in pure flavor states, so ${\bf s}$ and ${ \bar{\bf s}}$ initially point in the $\hat{\bf z}$ and $-\hat{\bf z}$ directions.
\begin{figure}[thb]%
  \subfigure[tight][Antineutrinos initially dominate.]{\label{fig:adiabatic}\includegraphics[width=0.45\textwidth,trim=0 0 0 0,clip]{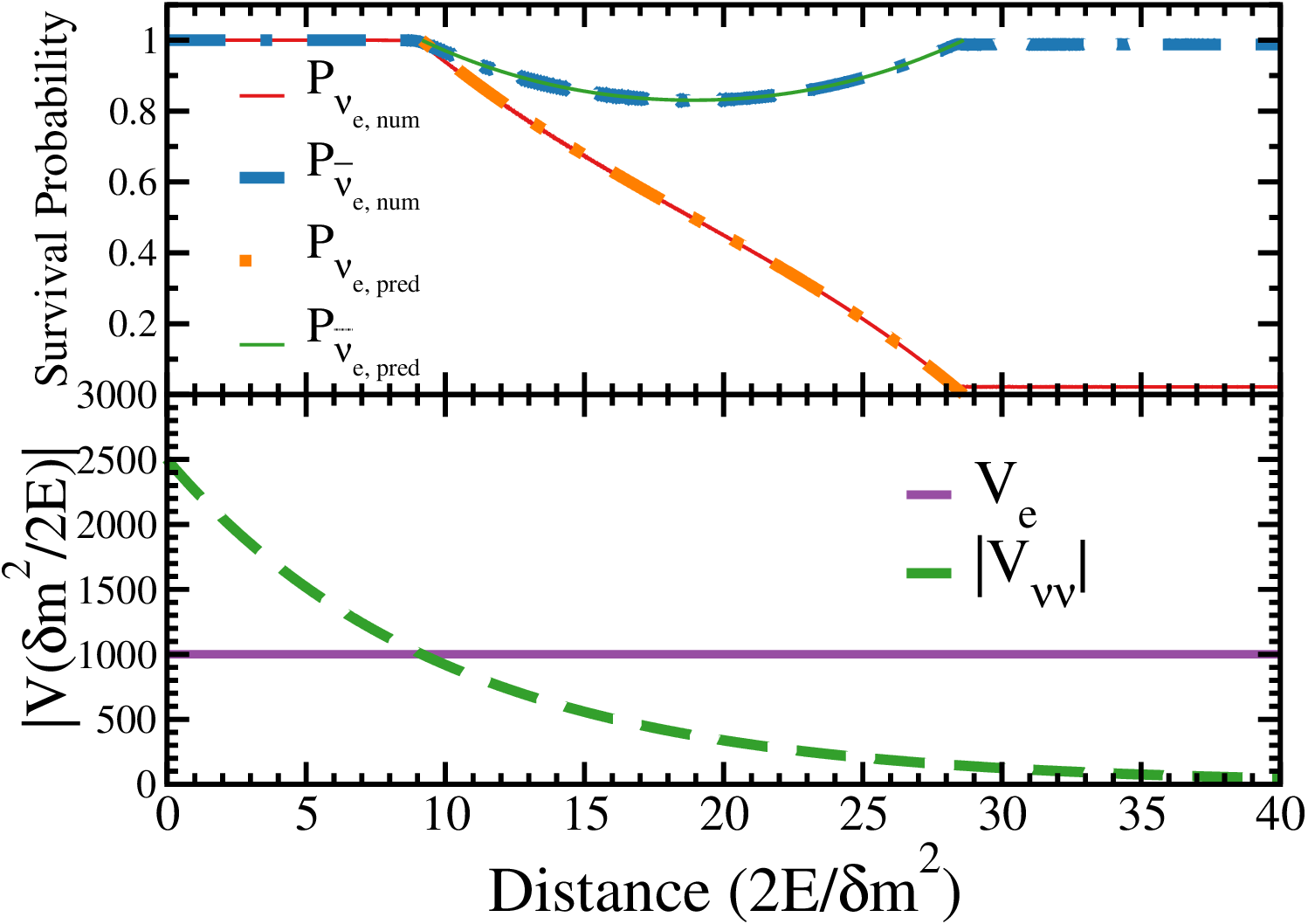}}\hfill%
  \subfigure[tight][Matter initially dominates.]{\label{fig:diabatic}\includegraphics[width=0.45\textwidth,trim=0 0 0 0,clip]{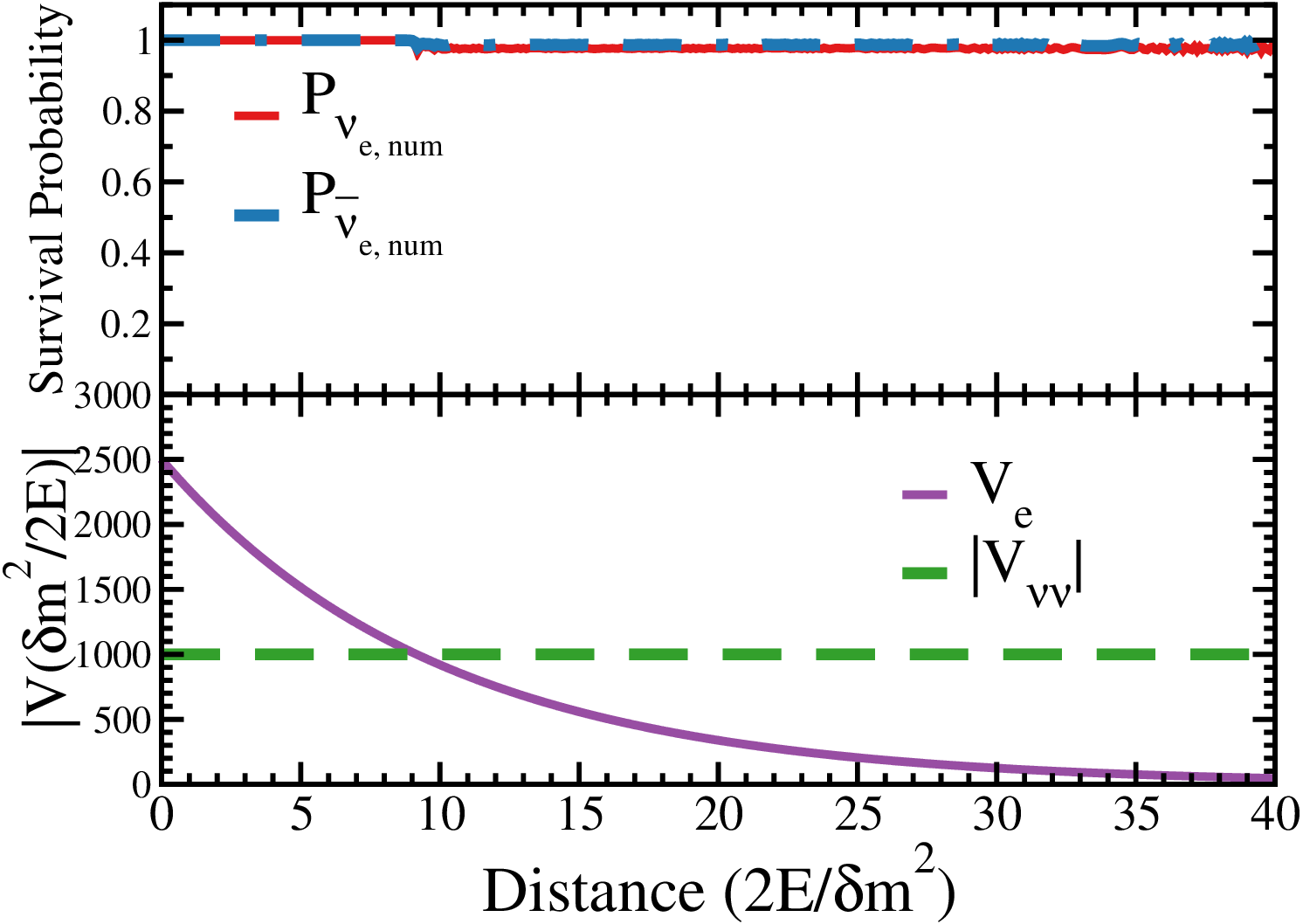}}%
  \raggedright{}
  \vspace{-15pt}
 \caption{{\em Top panel, both plots}: Survival probabilities $P_{\nu_e}$ (solid red line) and $P_{\bar{\nu}_e}$ (dashed blue line).
   {\em Bottom panel, both plots}: Potentials in units of $\Delta$. 
   The purple solid line shows the magnitude of the neutrino-electron potential, $V_e(l)$, 
   and the green dashed line shows the unoscillated neutrino-neutrino interaction potential, $|V_{\nu \nu}| = \mu(l)(\alpha - 1)$.}%
 \label{fig:diabaticandandadiabaticExample}%
\end{figure}

We perform two types of calculations with this single-energy configuration.
In the first case, we begin with $V_e(l=0) < \left|\mu_\nu (l=0)\left(1 - \alpha \right)\right|$ so that the neutrino self-interaction potential is initially greater than the matter potential.
We then allow $\mu_\nu$ to decline exponentially, while keeping $V_{e}$ constant, so that we can pass through the region where $V_e(l) =  \left|\mu_\nu(l)\left(1 - \alpha \right)\right|$.
In the second case, we instead start with $V_e(l=0) > \left|\mu_\nu (l=0) \left(1 - \alpha \right)\right|$ corresponding to the situation where the matter potential initially dominates.
We then keep $\mu$ fixed and allow $V_e(l)$ to decline so that, once again, at some point $V_e(l) = \left|\mu_\nu(l) \left(1 - \alpha \right)\right|$.
We demonstrate the results of this calculation in Fig.\ \ref{fig:diabaticandandadiabaticExample} with the specific functional forms $V_e(l) =1000\Delta$, $\mu_\nu(l)(\alpha-1)  = 10000\Delta e^{\left(-\frac{ \Delta}{10}l\right)} $ in Fig.\ \ref{fig:adiabatic}, and $V_e(l) = 10000\Delta e^{\left(-\frac{ \Delta}{10} l\right)}$, $\mu_\nu (\alpha-1) =1000 \Delta$,  in Fig.\ \ref{fig:diabatic}. In both Figs., $\alpha =4/3$. 
The top panels of these plots show that the scenario where $\mu_\nu(l) \left(\alpha - 1 \right)$ initially dominates over $V_e(l)$ produces a transition, while the reverse scenario does not.

Observe that the transition in  Fig.~\ref{fig:adiabatic} takes place over an extended period of time. 
The form of the potentials determines how long the system takes to go from the beginning, $V_e(l_i)  \approx  \mu_\nu(l_i)\left(\alpha - 1 \right)$ to the end, $ V_e(l_f)  \approx  \mu_\nu(l_f)\left(1 + \alpha \right)$. 
The duration of the transition is $\delta l_{1} \sim \tau_{V_e / \mu} \ln ((1 + \alpha) / (\alpha - 1))$,
where $\tau_{V_e / \mu}$ is the effective scale height of the ratio of the matter potential to the neutrino potential,, $\tau_{V_{e}/\mu}=|d \ln (V_{e}/\mu_\nu)/dl|)^{-1}$.
During this time, the system \emph{maintains a position approximately on the resonance}, i.e. $V_z(l) \approx V_e(l) +  \mu_\nu(l)\left(s_z + \alpha \bar{s}_z \right)$ hovers around zero. 
Both ${\bf s}$ and ${\bf \bar s}$ transform to maintain a cancellation between the self-interaction and the matter terms.
This behavior differs both from standard MSW \cite{Lunardini:2003eh,Dighe:1999bi} where the system passes quickly through the place where $V_z(l)\simeq0$ and also from synchronized oscillation where the neutrinos and antineutrinos are locked.

The transition behavior can be described analytically.
Examining the sum of Eqs.\ (\ref{eq:NFIS1}) and (\ref{eq:NFIS2}) as well as the behavior in Fig.\ \ref{fig:adiabatic}, we see that precession 
around the z-axis is nearly absent so that during the transition ${\bf s} +  \alpha {\bf \bar{s}}$ grows along the z-axis only.
By combining $s_x \approx - \alpha\bar{s}_x$, $s_y  \approx -\alpha\bar{s}_y$, and $V_z(l) \approx 0$, along with the approximation $\Delta\cos2\theta_{V} \approx 0$ we find
\begin{align}
  s_z &\approx \frac{\left(\alpha^2 - 1\right) \mu_\nu(l)^2 - V_e(l)^2}{4 V_e(l) \mu_\nu(l)},
  \label{eqn:foundPz}\\
  \bar{s}_z&\approx  -  \frac{\left(\alpha ^2 -  1\right) \mu_\nu(l)^2 + V_e (l)^2}{4 \alpha  V_e(l) \mu_\nu(l)}. \label{eqn:foundPzBar}
\end{align}

In Fig.~\ref{fig:adiabatic}, starting at the initial resonance point, we plot our analytic estimate of the
survival probabilities from Eqs.\ (\ref{eqn:foundPz}) and (\ref{eqn:foundPzBar}), using $P_{\nu_e} = s_z + 1/2$ and $P_{\bar{\nu}_e} = -\bar{s}_z + 1/2$. The agreement with the numerical evolution is evident.
If we try the same for Fig.\ \ref{fig:diabatic}, we do not find allowed solutions for the survival probability, in accord with the figure. 

It can be further seen that an initially dominant self-interaction potential is not, in general, sufficient to induce an MNR transition.
The mixing angle, $\theta_V$ also plays a role. 
Indeed, the vacuum term $\Delta\sin2\theta_{V}$ is the only physical source of flavor violation in this system.
From inspection of the sum of Eqs.\ (\ref{eq:NFIS1}) and (\ref{eq:NFIS2}), we see that the distance scale of the transition is $\delta l_2 \approx \alpha / ( \Delta \sin 2 \theta_V \langle s_y - \alpha\bar{s}_y \rangle)$, where $\langle \cdot \rangle$ is the average value during the transition.
For the scales $\delta l_1$ and $\delta l_2$ to be compatible, $\langle s_y - \alpha\bar{s}_y \rangle$ must adjust to $\theta_V$ and $\tau_{V_e / \mu}$, but for sufficiently small $\Delta \sin 2 \theta_V$ this condition cannot be fulfilled and hence the MNR transition is not realized.
In the example in Fig. \ref{fig:adiabatic}, if the mixing angle is reduced by an order of magnitude or more, then little transition occurs. 

We note that the MNR transformation does not depend on the sign of the mass hierarchy, as long as the matter and self-interaction potentials stay well above the vacuum scale.
However, it does depend on the asymmetry between electron neutrinos and antineutrinos.  The fluxes of the non-electron flavors ($\nu_\mu$, $\nu_\tau$ and their antiparticles) also play a role, as discussed next.

\begin{figure}[htb]
  \raggedright{}%
  \subfigure[tight][$\beta=0.7$]{\label{fig:muon0.2}\includegraphics[width=0.45\textwidth,trim=0 0 0 0,clip]{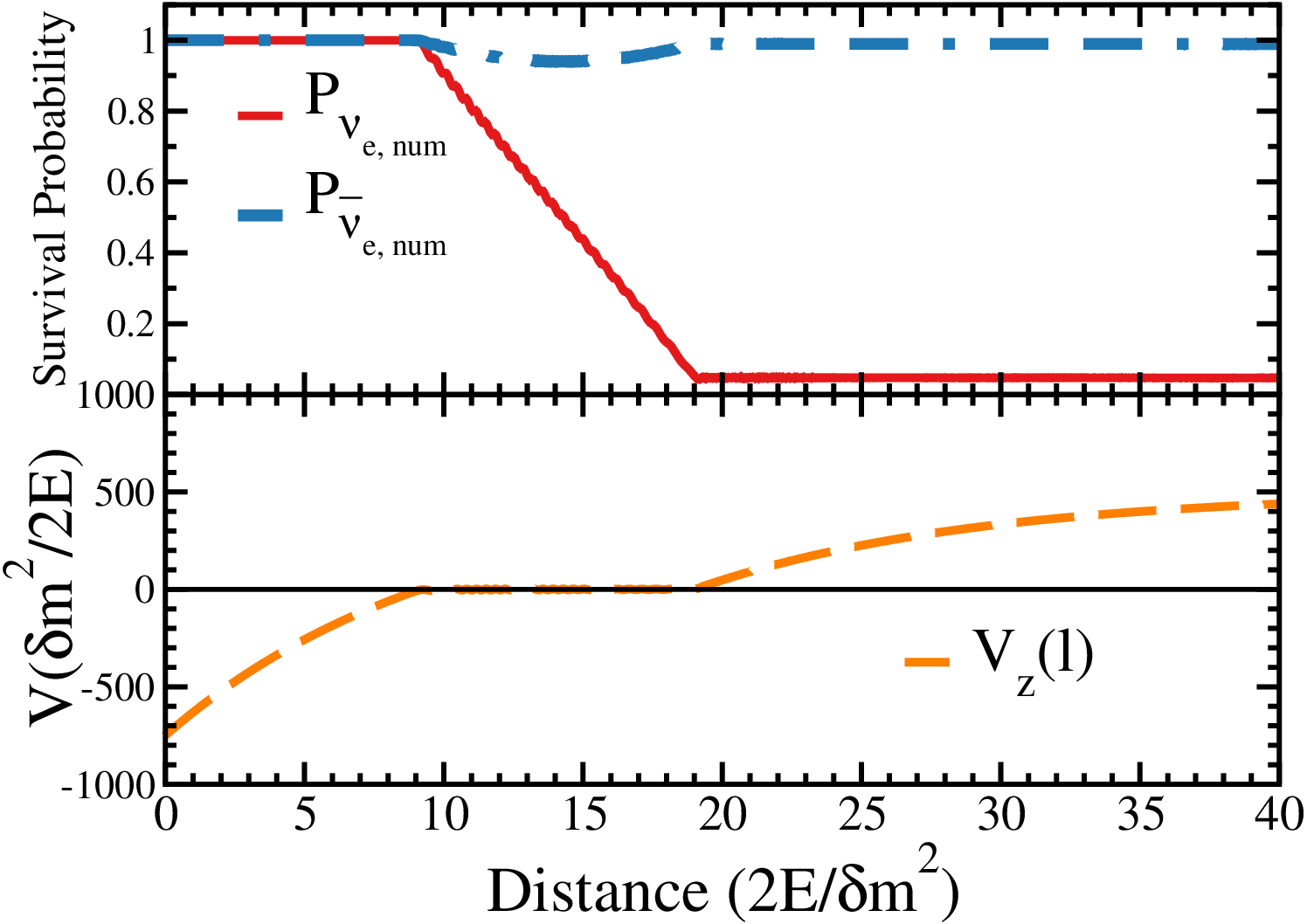}}\hfill%
  \subfigure[tight][$\beta=1.3$]{\label{fig:muon1.15}\includegraphics[width=0.45\textwidth,trim=0 0 0 0,clip]{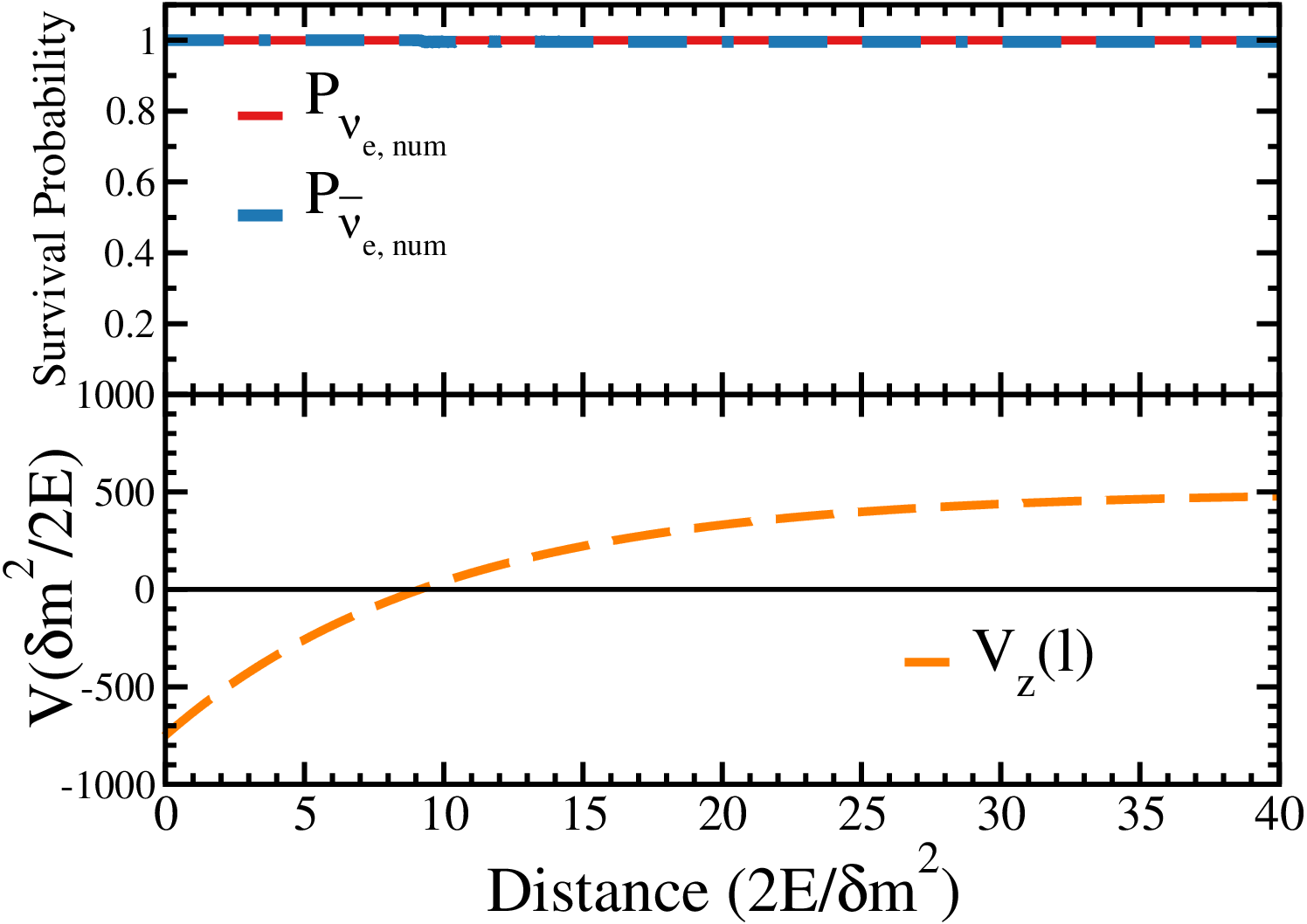}}
  \vspace{-15pt}
  \caption{
    {\em Top panel, both plots}: Survival probabilities, $P_{\nu_e}$ (solid red line) and $P_{\bar{\nu}_e}$ (dashed blue line).
    {\em Bottom panel, both plots}: Potentials $V_z(l)$.
    Fig.\ \ref{fig:muon0.2} shows $\beta=0.7$ and Fig.\ \ref{fig:muon1.15} shows $\beta=1.3$, where $\beta$ is the ratio of the muon flux to the electron neutrino flux.
   In both cases the overall potential begins dominated by the neutrino-neutrino interaction term, $V_z(l)<0$.
  }\label{fig:muonNeutrinos}
\end{figure}

{\it Suppression of the matter-neutrino resonance transition from $\nu_\mu$ and $\nu_\tau$:}
Disks from compact object mergers will not only emit electron neutrinos and antineutrinos, but also $\nu_\mu$, $\nu_\tau$, $\bar{\nu}_\mu$, and  $\bar{\nu}_\tau$.
We explore the importance of $\nu_\mu$, $\nu_\tau$ to the matter neutrino resonance transition, by considering four types of neutrinos, $\nu_e$, $\bar{\nu}_e$, $\nu_\mu$ and $\bar{\nu}_\mu$, all with the same energy.
In Fig.\ \ref{fig:muonNeutrinos} we have taken the model used to make Fig.\ \ref{fig:adiabatic}, and added to it muon neutrino and antineutrino fluxes of the same energy.
In Fig.\ \ref{fig:muonNeutrinos}, the fluxes of the muon neutrino and antineutrino are equal and their ratio relative to the electron neutrino flux is $\beta$.  We see from these figures that a sufficiently large $\nu_\mu$ flux suppresses the transition.
The bottom panels show that when the system exhibits a transition it maintains the resonance $V_z(l) \approx 0$, but when it does not, $V_z(l)$ passes straight through zero.

Net flavor isospin vectors, e.g. $\mu (l) s_z = \mu_e ( s_{\nu_e,z} + \beta s_{\nu_\mu,z})$ are helpful in understanding the behavior in Fig.\ \ref{fig:muonNeutrinos}. 
The net vector is reduced as the flux of muon neutrinos increases and eventually switches sign.
Since transition depends on the ability of the NFIS vectors to rotate in such a way that $V_z(l) \approx 0$ is maintained and the $x$ and $y$ components cancel, if the net vector is reduced to almost zero, then this becomes impossible.
Therefore, a muon neutrino flux comparable to the electron neutrino flux suppresses the transition.
Similarly if the muon antineutrino and electron antineutrino fluxes are comparable then the transition is suppressed.

{\it Merger Disk Calculations:} Determining how many neutrinos are emitted from a compact object merger disk is clearly a complex task.
The emission produces an energy spectrum for all types neutrinos $\nu_e$, $\nu_\mu$, $\nu_\tau$, $\bar{\nu}_e$, $\bar{\nu}_\mu$, and $\bar{\nu}_\tau$.
While different predictions for the flux and energy distributions of $\nu_e$ and $\bar{\nu}_e$ are typically in agreement, the fluxes of non-electron neutrinos are less certain.
Estimates of the $\nu_{\mu,\tau}$ fluxes range from comparable to the flux of $\nu_e$ to a small fraction $\sim 20\%$ to $\sim 30\%$ \cite{Deaton:2013sla, Kiuchi:2012mk}. 

To perform calculations that address the multi-energy nature of the emitted flux, the complex geometry of the disk, and the emission of all types of neutrinos, we first need to determine representative conditions.
Guided by the results of compact object merger neutrino surface calculations, e.g. \cite{Caballero:2009ww}, we construct a disk with the same qualitative features, i.e. the disk emits all types of neutrinos with energy hierarchy, $E_{\nu_e}<E_{\bar{\nu}_e}<E_{\nu_{\mu,\tau}}$ and the number flux of $\bar{\nu}_e$ is largest, followed by $\nu_e$ and then $\nu_{\mu,\tau}$.
We choose the disk radius to be $R_0=4.5\times10^6$ cm and temperatures $T_{\nu_e}=6.4$ MeV, $T_{\bar{\nu}_e}=7.1$ MeV, and $T_{\bar{\nu}_{\mu,\tau}}=T_{\nu_{\mu,\tau}}=7.4$ MeV.
We assume that neutrinos are not emitted from the last stable orbit, as determined from a 3$M_\odot$ black hole at the center.  The disk size is expected to be smaller for the $\nu_\mu$ and $\nu_\tau$ than for $\nu_e$.  For ease of computation, we use the same disk size for each flavor of neutrino and take account of the smaller $\nu_\mu$, $\bar{\nu}_\mu$, $\nu_\tau$, $\bar{\nu}_\tau$ fluxes by scaling these fluxes relative to their blackbody values.

We generalize the calculation in the previous section to multi-energies and three flavors of neutrinos.
Our calculation technique is described in \cite{Galais:2011jh,Kneller:2009vd}.
We assume that the neutrinos are initially purely in flavor states, use the \lq\lq single angle'' approximation \cite{Dasgupta:2008cu}, and take the vacuum parameters to be $\delta m^2_{12}=7.6\times10^{-5}$eV$^2$, $\delta m^2_{32}=-2.4\times10^{-3}$eV$^2$,	
$\theta_{12}=0.60$, 
$\theta_{13}=0.16$ and 
$\theta_{32}=0.76$, 
which are values consistent with the Particle Data Group's favored parameters \cite{Beringer:1900zz}.

\begin{figure}[b]
  \subfigure[Fluxes of $\nu_\mu$, $\nu_\tau$, $\bar{\nu}_\mu$, and $\bar{\nu}_\tau$ are scaled so that the emitted number flux is $35\%$ of blackbody.]{\label{thirtyFivePercent}\includegraphics[trim=0 0 0 0,clip,width=0.45\textwidth]{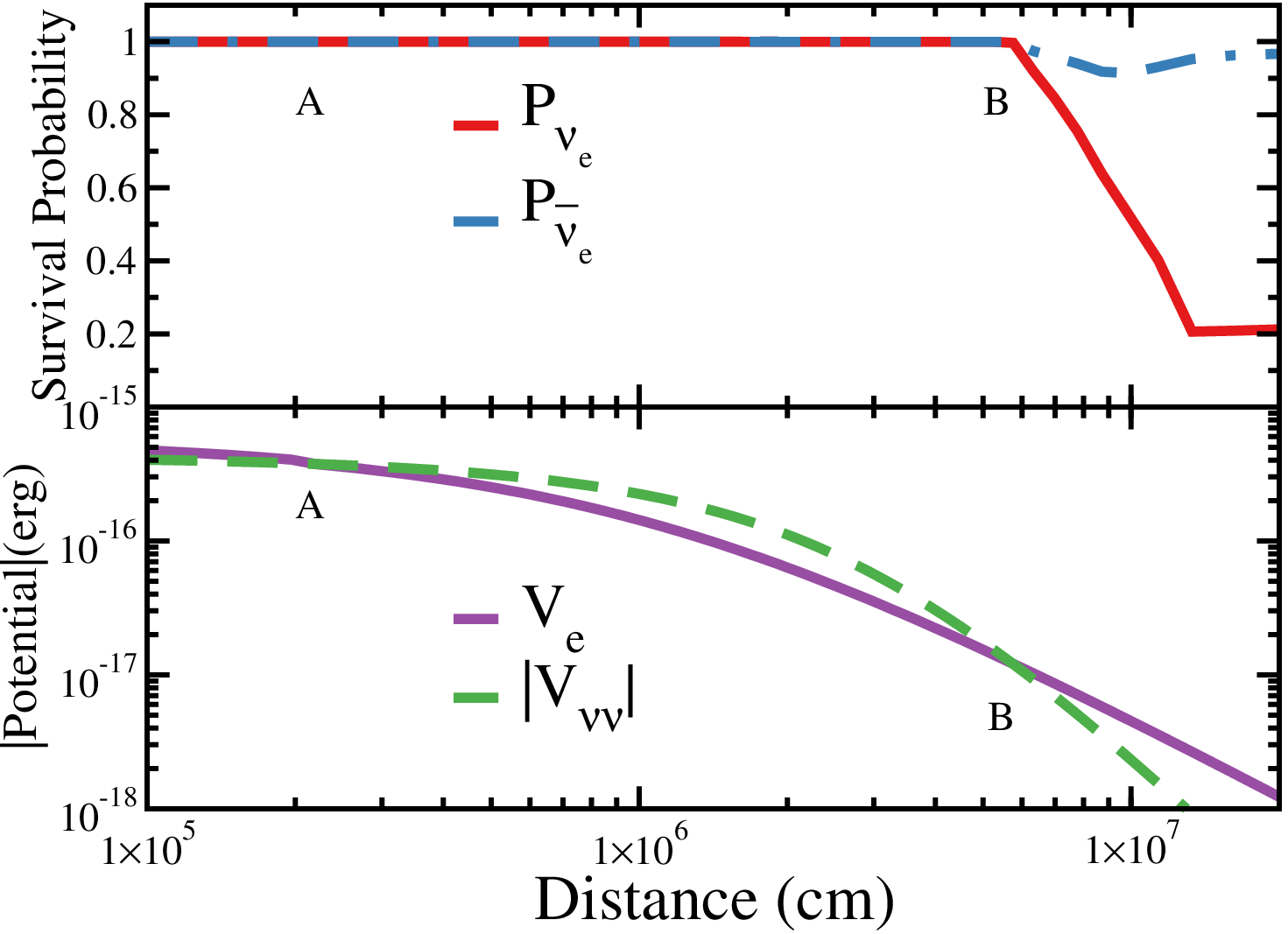}}\\
  \subfigure[tight][Fluxes of $\nu_\mu$, $\nu_\tau$, $\bar{\nu}_\mu$, and $\bar{\nu}_\tau$ are scaled so that the emitted number flux is $40\%$ of blackbody.]{\label{fortyPercent}\includegraphics[trim=0 0 0 0,clip,width=0.45\textwidth]{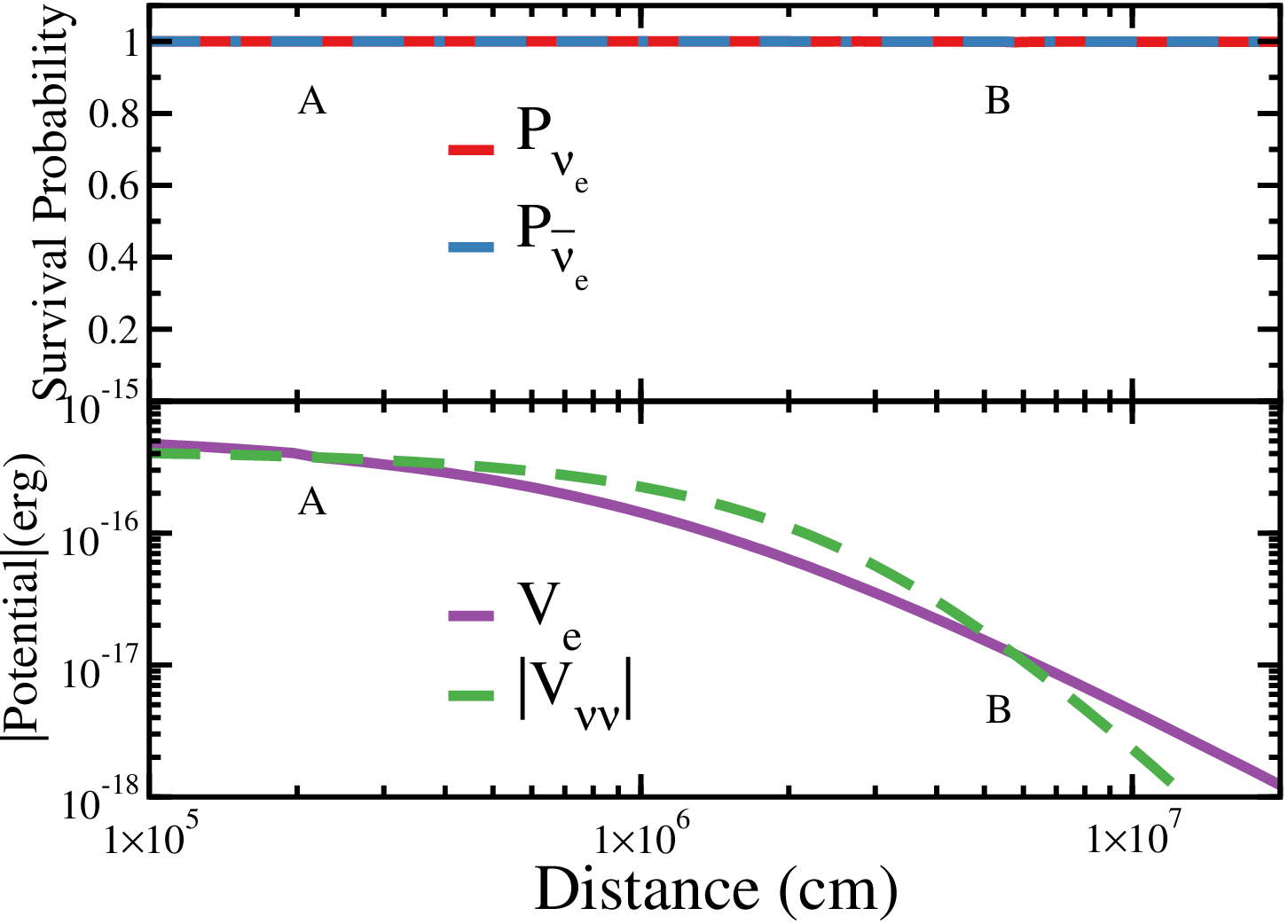}}
 \caption{ 
     {\em Top panel, both plots}: 
     The solid red (dashed blue) line shows the survival probability for an electron neutrino (antineutrino) as a function of progress along the trajectory.
     {\em Bottom panel, both plots}:
     The solid purple line shows the matter potential, $V_e(r)$
     The dashed green line shows the magnitude of the neutrino self interaction potential, $|V_{\nu \nu}|$ in the absence of oscillation.
     Crossing (resonance) points and are indicated by the letters A and B on the plot.
 }\label{lowDensityWithHeavy}
\end{figure}

We report results in Fig.\ \ref{lowDensityWithHeavy} 
for a neutrino moving along the same trajectory that might be taken by an outflowing mass element \cite{Surman:2004sy}, which begins at an initial disk radius of $r_0=2.2\times10^6$ cm. While the material lifts initially vertically from the disk, it later takes a radial trajectory.
Since we are {\it not} considering a trajectory emitted vertically above the black hole, we {\it cannot} rely on the disk symmetry to simplify the calculation.
Instead, we use the geometric factor that describes the decline of the neutrino fluxes as a function of distance from the disk from \cite{Malkus:2012ts}. The top panels in Fig.~\ref{thirtyFivePercent} and Fig.~\ref{fortyPercent} show the energy integrated survival probability.  In the bottom panels of each figure, we show the overall relative strengths of each part of the potential, the matter potential $V_e(r)$ and the unoscillated neutrino self interaction potential $|V_{\nu \nu} (r)|$.

The results depicted in Fig. \ref{thirtyFivePercent}  confirm that the MNR transition occurs as predicted.
We see that the crossing points A and B produce different behavior.
A careful examination of the bottom panel of Fig.\ \ref{thirtyFivePercent}, shows that at crossing point A, the system begins matter dominated, while at crossing point B, it begins neutrino dominated.
Consistent with the behavior of the single energy calculation, point A produces no transition, while point B produces a neutrino matter resonance transition.
For situations like Fig.\ \ref{thirtyFivePercent}, where the mu/tau contribution is small one can apply the timescale arguments from the single energy model.
The asymmetry is $\alpha = 1.37$ and the potential ratio scale height is $\tau_{V_e/\mu} = 5.8\times10^6$cm,
so the system should exhibit the MNR transition for $\theta \gtrsim 2.3 \times 10^{-2}$, which is safely fulfilled by the measured value of $\theta_{13}$ \cite{An:2013zwz}.
Again consistent with the single energy calculation, from a comparison of Fig.\ \ref{thirtyFivePercent}, with Fig.\ \ref{fortyPercent}, we see that there is an abrupt change in the transition behavior when the $\mu$ and $\tau$ type neutrino fluxes become larger than a certain size. 

{\it Conclusions:} We find a novel mechanism of collective neutrino flavor transformations, MNR, which operates in the compact object merger disk environments, when the initially dominant neutrino self-interaction potential becomes equal to the matter potential. The phenomenon owes its existence to the large measured value of $\theta_{13}$ and occurs for both types of the neutrino mass hierarchy. 
The transition behavior is not finely tuned; it occurs over a wide range of disk radii, densities and neutrino fluxes. The transition does depend on the size of the asymmetry between $\nu_{e}$ and $\bar\nu_{e}$, as well as on the  $\nu_\mu$ and $\nu_\tau$ fluxes (currently predicted to be relatively small \cite{Deaton:2013sla,Kiuchi:2012mk}). 
 
Importantly, MNR transitions occur relatively close to the surface of the disk 
and hence may influence wind type nucleosynthesis, such as r-process \cite{Surman:2008qf} or nickel production \cite{Surman:2013sya}. The MNR phenomenon, therefore, may have
observable consequences in the galactic inventory of elements, or in the electromagnetic signal from mergers, sometimes called a kilonova \cite{Metzger:2011bv}. These signatures need to be modeled in future work.

Further work should also include improved modeling of the neutrino fluxes of all flavors in the compact object merger environments, as well as more accurate neutrino flavor transformation physics, 
for example, including multi-angle \cite{Duan:2006jv,Duan:2010bf,Mirizzi:2013rla,Raffelt:2013rqa} and halo effects \cite{Cherry:2013mv}.

We thank John F.\ Cherry, Jim Kneller, Rebecca Surman and Cristina Volpe for useful discussion.
This work was supported in part by U.S. DOE Grants No. DE-FG02-02ER41216 and DE-SC0004786, and in part by the LANL LDRD program.
GCM and AF would like to thank CETUP* (Center for Theoretical Underground Physics and Related Areas), 
for its hospitality and partial support during the 2013 Summer Program.


\end{document}